# A REQUIEM FOR SCHRÖDINGER'S CAT

THOMAS L. WILSON
*National Aeronautics and Space Administration*
*Houston, TX USA 77058*
*Thomas.Wilson@cern.ch*

Schrödinger introduced his celebrated "cat" paradox[1] in 1935 in response to the claim by Einstein, Podolsky, & Rosen (EPR)[2] that quantum mechanics was incomplete. EPR had presented the possibility that pairs of particles can be strongly correlated over large space-like distances. Schrödinger coined the term entanglement for this quantum non-classical nonlocal correlation, which he claimed was an essential characteristic of quantum mechanics. The superposition principle in quantum mechanics allows for this, whereby the preparation of microscopic quantum systems into mixed states can be created by combinations of pure individual states. This has been corroborated in numerous experiments[3].

The physics question raised is whether quantum superposition can result in entangled macroscopic states (ones that can be seen by the aided human eye). That is, how universal is the linear superposition principle in macroscopic physics knowing that it is based mathematically upon the linearity of Hilbert space[4] in quantum physics. If $|1\rangle$ and $|2\rangle$ are two pure states or rays in Hilbert space, quantum mechanics requires that any normed linear combination $\alpha|1\rangle + \beta|2\rangle$ also represents a possible state.

Schrödinger devised a cat as such a macroscopic state meant to illustrate the absurdity in the EPR paradox. This cat paradox is legion and depicts the conflict between the quantum world which cannot be seen with the naked eye and the everyday classical world of experience which can. Can decoherent macroscopic objects like a cat be represented as rays in a quantum Hilbert space where the cat is both dead and alive at the same time?

We intend to show here that there is a serious error in Schrödinger's reasoning, and conclude that there is no paradox. All of the quantum physics[3,4] will nevertheless remain intact.

Schrödinger depicted the following experiment. A cat is placed in a sealed box along with a poisonous flask and a radioactive source. The box is sealed against environmentally-induced quantum decoherence, and contains a Geiger counter. If the counter detects radiation, the flask is broken and the poison kills the cat. The Copenhagen interpretation of quantum mechanics[5] implies that after a period in the box before the flask is broken, the cat is both dead and alive. Schrödinger (and others) claimed this possibility is ludicrous.

The fallacy in Schrödinger's cat paradox lies in the definition of life, our understanding of which has radically changed since his cat was introduced. There is no rigorous definition of life today except that it is a process. A living thing procreates, shares a common ancestor, is a cell or collection of cells, and contains DNA and/or RNA. Now, let's identify things that are not alive. Complex chemicals are capable of duplication, replication, and self-assembly[6]. But these properties are not the same as life. A virus is not alive; rather it is a complex protein that hijacks cells which then duplicate it to infect more cells. DNA and RNA are not alive but are coded recipes for proteins used in the design of living things. Cat and human hair and nails are keratin, a family of fibrous structural proteins that make up the outer layer of skin and are the key structural component of horns, hooves and other tissue in reptiles, birds, amphibians, and mammals. But hair, in particular cat hair, is not alive. Another example is chitin, a long-chain polymer that is the main component of cell walls in fungi, the exoskeletons of arthropods like crustaceans and insects. Chitin, however, is also not alive.

The list goes on of physical structures that are constituents of living things but are inanimate and not alive. Since all living things must contain DNA (not alive), then all living things are both dead and alive. Hence, Schrödinger's cat is both dead and alive (D&A) to begin with. It enters the experiment |D&A> and it exits the experiment |D&A> or |D>. To quote from Schrödinger's very own paper (Ref. 1, §10), *voilà tout* (there you have it). There is no paradox.

The question of what is life is one of the most important in science today. The very fact that Schrödinger paid so much attention to it[7] demonstrates that it is a significant question in physics. The subject is new: living matter physics treating living matter as another state or phase transition of matter, as opposed to condensed matter physics. Its answer should appeal to the current generation of new physicists. Until physicists, biologists, and biochemists can define the physical nature of living matter, science in general will have to wait for an ultimate answer to this profound question.